\documentclass[aps,showpacs,superscriptaddress,preprint]{revtex4}%
\usepackage{amsfonts}
\usepackage{amsmath}
\usepackage{amssymb}
\usepackage{graphicx}%
\begin{document}

\title{Chiral topological excitonic insulator in semiconductor quantum wells}
\author{Ningning Hao}
\affiliation{Institute of Physics, Chinese Academy of Sciences, Beijing 100190, People's
Republic of China}
\affiliation{LCP, Institute of Applied Physics and Computational Mathematics, P.O. Box
8009, Beijing 100088, People's Republic of China}
\author{Ping Zhang}
\thanks{Corresponding author; zhang\_ping@iapcm.ac.cn}
\affiliation{LCP, Institute of Applied Physics and Computational Mathematics, P.O. Box
8009, Beijing 100088, People's Republic of China}
\affiliation{Center for Applied Physics and Technology, Peking University, Beijing 100871,
People's Republic of China}
\author{Jian Li}
\affiliation{Institute of Physics, Chinese Academy of Sciences, Beijing 100190, People's
Republic of China}
\author{Zhigang Wang}
\affiliation{LCP, Institute of Applied Physics and Computational Mathematics, P.O. Box
8009, Beijing 100088, People's Republic of China}
\author{Wei Zhang}
\affiliation{LCP, Institute of Applied Physics and Computational Mathematics, P.O. Box
8009, Beijing 100088, People's Republic of China}
\author{Yupeng Wang}
\affiliation{Institute of Physics, Chinese Academy of Sciences, Beijing 100190, People's
Republic of China}


\begin{abstract}
We present a scheme to realize the chiral topological excitonic insulator in
semiconductor heterostructures which can be experimentally fabricated with a
coupled quantum well adjacent to two ferromagnetic insulating films. The
different mean-field chiral topological orders, which are due to the change in
the directions of the magnetization of the ferromagnetic films, can be
characterized by the TKNN numbers in the bulk system as well as by the winding
numbers of the gapless states in the edged system. Furthermore, we propose an
experimental scheme to detect the emergence of the chiral gapless edge state
and distinguish different chiral topological orders by measuring the thermal conductance.

\end{abstract}
\maketitle

\section{\text{Introduction}}

The search for new phases of quantum matter is one of the essential topics in
condensed-matter physics. Chiral topological band insulators (TBIs) are such a
type that has been attracting a lot of interest both theoretically and
experimentally. Although like trivial insulators in the sense that TBIs have a
band gap in the bulk, they are fundamentally distinguished from trivial ones
by their having gapless modes on the boundaries. These gapless modes are
robust under perturbations and cannot be gapped without going through a
quantum phase transition. In the case of time reversal symmetry (TRS)
breaking, a well-known TBI system is the Haldane's model which is a minimal
model to illustrate quantum anomalous Hall effect (QAHE) \cite{Haldane}. The
QAHE topological phase is characterized by the (Thouless, Kohmoto,
Nightingale, and Nijs) TKNN number \cite{TKNN} of the first Chern class of a
U(1) principal fiber bundle on a torus in the bulk system or the winding
number of Halperin's edge-state theory \cite{Halperin,Hatsugai} on the
boundary of the system. The coherence of the two different kinds of numbers is
guaranteed by the bulk-edge correspondence. Since the rigorously prerequisite
magnetic field in Haldane's model is difficult to realize in experiment,
recently, there are some new proposals \cite{Liu} to realize QAHE based on
single-particle picture.

In analogy with QAHE in single-particle picture, the superconductors in
TRS-broken ($p_{x}+ip_{y}$) weak pairing state in two dimensions with a fully
stable bulk gap opened by electron-electron interaction can also have chiral
topological order \cite{Read}. The edge states of the chiral superconductor
have half of the degrees of freedom compared to QAHE states due to the
particle-hole symmetry (PHS) and are called Majorana edge sates. In the spirit
of analogy with superconductor, a natural and important issue is how to get
chiral topological excitonic insulator (TEI), which is addressed in this
paper. \begin{figure}[ptb]
\begin{center}
\includegraphics[width=0.8\linewidth]{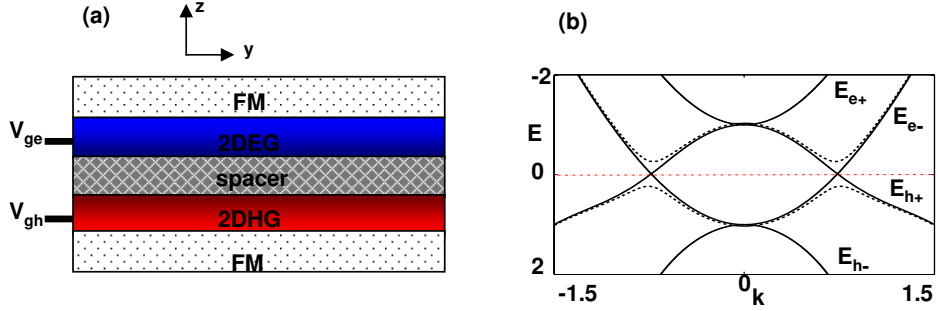}
\end{center}
\caption{(Color online) (a) Schematic structure of semiconductor quantum-wells
system that holds chiral TEI. The external gates (V$_{ge(h)}$) can
independently tune the chemical potential $\mu_{e(h)}$ to obtain the electron
and heavy-hole layer. The ferromagnetic insulating films support effective
exchange fields ($V_{e},V_{h}$). (b) The energy spectrum of the electron/hole
bilayer system near the Fermi energy $E_{F}\approx0$. Here the solid lines
denote non-interacting single-particle energy spectrum $E_{e(h)\pm}$, while
the dashed lines denote the exciton energy spectrum with an obvious mean-field
gap opened. We take $t_{e}=t_{h}=1$, $\mu_{e}=\mu_{h}=-4$, $(V_{e}%
,V_{h})=(1,1)$, $\alpha=0.5$, and $\beta=0.5$.}%
\end{figure}

In this paper, we consider an independently gated double-quantum-well
structure separated by a spacer as shown in Fig. 1(a). The ferromagnetic
insulating films are introduced to break TRS by inducing an effective Zeeman
splitting in the two dimensional electron (hole) gas [2DE(H)G]. The
magnetization is perpendicular to the two-dimensional layer. Note that the
orbital effect of the ferromagnetic films to the 2DE(H)G can be neglected due
to the local exchange interaction on the interface. An electron-hole fluid is
created by modulating the voltages so that the Zeeman-split upper branch of
the heavy-hole bands in 2DHG layer can move above the Zeeman-split lower
branch of the electron bands in 2DEG layer. This procedure results in
spatially separated but strongly interacting electron and hole fluids if the
two layers are close enough. The external electric field produced by the bias
voltages and the intrinsic electric field due to doping in the process of
fabricating quantum wells can enhance the structural inversion asymmetry and
induce the tunable Rashba spin-orbit (SO) interaction
\cite{Nitta,Lu,Grundler,Papadakis}. Recall that due to the strong SO coupling
and non-centrosymmeytic property \cite{Gor'kov}, the broken parity of the
order parameter is the prerequisite condition of the chiral superconductor. In
analogy with chiral superconductor, we demonstrate that the chiral TEI can
occur when the Rashba SO interaction is strong enough with respect to the
amplitude of the excitonic order parameter (EOP). Generally the Rashba SO
interaction strength can be influenced by carrier density, gated voltages,
material of quantum wells, etc \cite{Engels,Winkler,Winkler1}. In InAs
heterostructures, for instance, this quantity can be electrically tuned to be
as large as $\alpha\mathtt{\approx}$50 meV \AA \ \cite{Yang}. Due to the
missing PHS, the edge states of chiral TEI in the present system are not
Majorana fermions. The implication of these gapless edge states for
experimental observations is also discussed in this paper.

\section{Model and Hamiltonian}

We start with an effective electron/hole semiconductor bilayer system confined
in the $x$-$y$ plane. Here for the hole layer only the heavy-hole bands are
occupied as in typical experiments, while the light-hole bands are empty and
are therefore not taken into account in our model. The resultant tight-binding
Hamiltonian for the Rashba spin-orbit coupled semiconductor bilayer system is
$\mathcal{H}$=$\sum_{p}(\mathcal{H}_{kin}^{(p)}$+$\mathcal{H}_{R}^{(p)}%
)$+$\mathcal{H}_{int}^{(e-h)}\mathtt{\equiv}\mathcal{H}_{0}$+$\mathcal{H}%
_{int}^{(e-h)}$:%
\begin{align}
\mathcal{H}_{kin}^{(p)}  &  =\underset{<i,j>,\sigma}{\sum}(-t_{p}-\mu
_{p}\delta_{ij})p_{i,\sigma}^{\dagger}p_{j,\sigma}+\underset{j,\sigma
,\sigma^{\prime}}{\sum}V_{p}\tau_{p}(s_{z})_{\sigma,\sigma^{\prime}%
}p_{j,\sigma}^{\dagger}p_{j,\sigma^{\prime}},\nonumber\\
\mathcal{H}_{R}^{(e)}  &  =\frac{1}{2}\alpha\lbrack\underset{j}{\sum
}(e_{j,\uparrow}^{\dagger}e_{j+\delta x,\downarrow}-e_{j,\uparrow}^{\dagger
}e_{j-\delta x,\downarrow})\nonumber\\
&  -i\underset{j}{\sum}(e_{j,\uparrow}^{\dagger}e_{j+\delta y,\downarrow
}-e_{j,\uparrow}^{\dagger}e_{j-\delta y,\downarrow})]+\text{H.c.},\nonumber\\
\mathcal{H}_{R}^{(h)}  &  =\frac{i}{2}\beta\lbrack\underset{j}{\sum
}(h_{j,\Uparrow}^{\dagger}h_{j+2\delta x,\Downarrow}-h_{j,\Uparrow}^{\dagger
}h_{j-2\delta x,\Downarrow})\nonumber\\
&  +i\underset{j}{\sum}(h_{j,\Uparrow}^{\dagger}h_{j+2\delta y,\Downarrow
}-h_{j,\Uparrow}^{\dagger}h_{j-2\delta y,\Downarrow})\nonumber\\
&  +3(1-i)\underset{j}{\sum}(h_{j,\Uparrow}^{\dagger}h_{j-\delta x+\delta
y,\Downarrow}-h_{j,\Uparrow}^{\dagger}h_{j+\delta x-\delta y,\Downarrow
})\label{Ham1}\\
&  +3(1+i)\underset{j}{\sum}(h_{j,\Uparrow}^{\dagger}h_{j-\delta x-\delta
y,\Downarrow}-h_{j,\Uparrow}^{\dagger}h_{j+\delta x+\delta y,\Downarrow
})\nonumber\\
&  +4\underset{j}{\sum}(h_{j,\Uparrow}^{\dagger}h_{j+\delta x,\Downarrow
}-h_{j,\Uparrow}^{\dagger}h_{j-\delta x,\Downarrow})\nonumber\\
&  +4i\underset{j}{\sum}(h_{j,\Uparrow}^{\dagger}h_{j+\delta y,\Downarrow
}-h_{j,\Uparrow}^{\dagger}h_{j-\delta y,\Downarrow})]+\text{H.c.},\nonumber\\
\mathcal{H}_{int}^{(e-h)}  &  =-\frac{1}{2}\underset{i,j,\sigma,\sigma
^{\prime}}{%
%TCIMACRO{\dsum }%
%BeginExpansion
{\displaystyle\sum}
%EndExpansion
}U_{i,j}^{(eh)}(d)e_{i\mathbf{\sigma}}^{\dagger}h_{j\mathbf{\sigma}^{\prime}%
}^{\dagger}h_{j\mathbf{\sigma}^{\prime}}e_{i\mathbf{\sigma}}\text{.}\nonumber
\end{align}
Here $t_{p}$ denotes the nearest-neighbor hopping amplitude while $\mu_{p}$
represents the chemical potential in electron ($p$=$e$) or heavy-hole ($p$%
=$h$) layer. $s_{z}$ is the $z$-component of the Pauli matrices and $V_{p}%
\tau_{p}$ is the effective Zeeman splitting ($\tau_{e}$=$1$ for electron layer
and $\tau_{h}$=$-1$ for hole layer). $p_{j,\sigma}$ is the fermion
annihilation operator at lattice site $j$ with spin $\pm$1/2 ($\uparrow
,\downarrow$) for $p$=$e$ and spin $\pm$3/2 ($\Uparrow,\Downarrow$) for
$p$=$h$. $\alpha$ ($\beta$) is the Rashba SO interaction strength in the
electron (heavy-hole) layer. $\delta x$ ($\delta y$) is the square lattice
spacing along the $x$ ($y$) direction. In the interaction term, $U_{i,j}%
^{(eh)}(d)$=$e^{2}/\varepsilon\sqrt{\left\vert \vec{R}_{i,e}-\vec{R}%
_{j,h}\right\vert ^{2}\text{+}d^{2}}$, where $\varepsilon$ is the dielectric
constant of the spacer and $d$ is the interlayer distance. We only consider
the interaction correlative to exciton formation and ignore the electron-hole
exchange interaction. The lattice Hamiltonian can be transformed into the
momentum space with the Fourier transformation $(e_{\vec{k},\sigma},h_{\vec
{k},\sigma})$=$1/\sqrt{\Omega}\sum_{i}e^{i\vec{k}\cdot\vec{R}_{i}}%
(e_{i,\sigma},h_{i,\sigma})$. The result reads%
\begin{align}
\mathcal{H}_{kin}^{(p)}(\vec{k})  &  =\underset{\vec{k},\sigma,\sigma^{\prime
},p}{%
%TCIMACRO{\dsum }%
%BeginExpansion
{\displaystyle\sum}
%EndExpansion
}[(\zeta_{\vec{k}}^{(p)}-\mu_{p})\delta_{\sigma,\sigma^{\prime}}\nonumber\\
&  +V_{p}\tau_{p}(s_{z})_{\sigma,\sigma^{\prime}}]p_{\vec{k},\sigma}^{\dagger
}p_{\vec{k},\sigma^{\prime}},\nonumber\\
\mathcal{H}_{R}^{(e)}(\vec{k})  &  =\underset{\vec{k}}{%
%TCIMACRO{\dsum }%
%BeginExpansion
{\displaystyle\sum}
%EndExpansion
}i\alpha(\sin k_{x}-i\sin k_{y})e_{\vec{k}\uparrow}^{\dagger}e_{\vec
{k}\downarrow}+\text{H.c.},\nonumber\\
\mathcal{H}_{R}^{(h)}(\vec{k})  &  =\underset{\vec{k}}{%
%TCIMACRO{\dsum }%
%BeginExpansion
{\displaystyle\sum}
%EndExpansion
}i\beta(a_{k}-ib_{k})h_{\vec{k}\Uparrow}^{\dagger}h_{\vec{k}\Downarrow
}+\text{H.c.},\label{Ham1_k}\\
\mathcal{H}_{int}^{(e-h)}  &  =-\frac{1}{2\Omega}\underset{\vec{k},\vec
{k}^{\prime},\vec{q},\sigma,\sigma^{\prime}}{%
%TCIMACRO{\dsum }%
%BeginExpansion
{\displaystyle\sum}
%EndExpansion
}U^{(eh)}(q)e_{\vec{k}\mathbf{+}\vec{q}\mathbf{\sigma}}^{\dagger}h_{\vec
{k}^{\prime}\mathbf{-}\vec{q}\mathbf{\sigma}^{\prime}}^{\dagger}h_{\vec
{k}^{\prime}\mathbf{\sigma}^{\prime}}e_{\vec{k}\mathbf{\sigma}},\nonumber
\end{align}
where $U^{(eh)}(q)$=$\frac{2\pi e^{2}}{\epsilon q}e^{-qd}$, $\zeta_{\vec{k}%
}^{(p)}$=$-2t_{p}(\cos k_{x}$+$\cos k_{y})$, $a_{k}$=$2(3\cos k_{y}\sin
k_{x}\mathtt{-}\sin k_{x}\cos k_{x}\mathbf{-}2\sin k_{x})$, and $b_{k}%
$=$2(-3\cos k_{x}\sin k_{y}$+$\sin k_{y}\cos k_{y}$+$2\sin k_{y})$. In the
above Hamiltonian, the interlayer tunneling is neglected, because the
insulating spacer can supply a high barrier to stop the direct interlayer
hopping. We also neglect the intralayer electron-electron and hole-hole
interactions, since they are expected to renormalize the single-particle
energy of each layer and have no essential influence on the topological
properties of the system. In the mean-field approximation, the above
Hamiltonian can be written as%
\begin{align}
\mathcal{H}_{MF}  &  =\underset{\vec{k},\sigma,\sigma^{\prime},p}{%
%TCIMACRO{\dsum }%
%BeginExpansion
{\displaystyle\sum}
%EndExpansion
}[(\zeta_{\vec{k}}^{(p)}-\mu_{p})\delta_{\sigma,\sigma^{\prime}}+V_{p}\tau
_{p}(s_{z})_{\sigma,\sigma^{\prime}}]p_{\vec{k},\sigma}^{\dagger}p_{\vec
{k},\sigma^{\prime}}\nonumber\\
&  +\underset{p}{%
%TCIMACRO{\dsum }%
%BeginExpansion
{\displaystyle\sum}
%EndExpansion
}\mathcal{H}_{R}^{(p)}-\frac{1}{2}\underset{\vec{k}\sigma\sigma^{\prime}}{%
%TCIMACRO{\dsum }%
%BeginExpansion
{\displaystyle\sum}
%EndExpansion
}(\Delta_{\sigma\sigma^{\prime}}(\vec{k})e_{\vec{k}\mathbf{\sigma}}^{\dag
}h_{-\vec{k}\mathbf{\sigma}^{\prime}}^{\dag}+\text{H.c.})\label{Ham1_mf}\\
&  +\frac{1}{2\Omega}\underset{\vec{k}\vec{q}\sigma\sigma^{\prime}}{%
%TCIMACRO{\dsum }%
%BeginExpansion
{\displaystyle\sum}
%EndExpansion
}\frac{\Delta_{\sigma\sigma^{\prime}}(\vec{k})\Delta_{\sigma\sigma^{\prime}%
}^{\ast}(\vec{k}-\vec{q})}{U^{(eh)}(q)},\nonumber
\end{align}
where the EOPs are defined as%
\begin{equation}
\Delta_{\sigma\sigma^{\prime}}(\vec{k})=\frac{1}{\Omega}\underset{\vec{q}}{%
%TCIMACRO{\dsum }%
%BeginExpansion
{\displaystyle\sum}
%EndExpansion
}U^{(eh)}(q)\left\langle h_{-\vec{k}+\vec{q}\mathbf{\sigma}^{\prime}}%
e_{\vec{k}-\vec{q}\mathbf{\sigma}}\right\rangle . \label{EOP0}%
\end{equation}
In the Nambu notation with combined $e$-$h$ field operator basis $\psi
$=$[e_{\vec{k}\mathbf{\uparrow}}$ $e_{\vec{k}\mathbf{\downarrow}}$
$h_{-\vec{k}\mathbf{\Uparrow}}^{\dagger}$ $h_{-\vec{k}\mathbf{\Downarrow}%
}^{\dagger}]^{T}$, the mean-field Hamiltonian is expressed as $\mathcal{H}%
_{MF}$=$\psi^{\dagger}H_{MF}\psi$+$const$ with%
\begin{equation}
H_{MF}=\left[
\begin{array}
[c]{cc}%
\mathbf{\Sigma}_{\vec{k}}^{(e)}-\mu_{e}+V_{e}s_{z} & \mathbf{\Delta}(\vec
{k})\\
\mathbf{\Delta}^{\dagger}(\vec{k}) & \mathbf{\Sigma}_{-\vec{k}}^{(h)}+\mu
_{h}+V_{h}s_{z}%
\end{array}
\right]  , \label{Ham1_mf1}%
\end{equation}
where $\mathbf{\Sigma}_{\pm\vec{k}}^{(p)}$=$\pm\zeta_{\pm\vec{k}}%
^{(p)}I\mathtt{\pm}\mathcal{H}_{R}^{(p)}$($\pm\vec{k}$) and
\begin{equation}
\mathbf{\Delta}(\vec{k})=-\frac{1}{2}\left[
\begin{array}
[c]{cc}%
\Delta_{\uparrow\Uparrow}(\vec{k}) & \Delta_{\uparrow\Downarrow}(\vec{k})\\
\Delta_{\downarrow\Uparrow}(\vec{k}) & \Delta_{\downarrow\Downarrow}(\vec{k})
\end{array}
\right]  \label{EOP1}%
\end{equation}
with $\Delta_{\sigma\sigma^{\prime}}(\vec{k})$ defined in Eq. (\ref{EOP0}).

The complex EOPs $\Delta_{\sigma\sigma^{\prime}}(\vec{k})$ can be
self-consistently obtained from exact numerical calculation of Eqs.
(\ref{Ham1_mf}) and (\ref{EOP0}) with respect to minimizing the ground state
energy. In our numerical calculation of $\Delta_{\sigma\sigma^{\prime}}%
(\vec{k})$, we set the lattice size 81$\times$81 and take $t_{e}$=$t_{h}$=$1$,
$\mu_{e}$=$\mu_{h}$=$-4$, $\alpha$=0.5, and $\beta$=$0$.5. There are four
different kinds of choices for the perpendicular magnetization in the two
magnetic films adjacent to the bilayer system. For the parallel
configurations, in our numerical simulations we choose $(V_{e},V_{h})$=$(1,1)$
and $(V_{e},V_{h})$=$(-1,-1)$, while for the antiparallel configurations we
choose $(V_{e},V_{h})$=$(-1,1)$ and $(V_{e},V_{h})$=$(1,-1)$. From our
extensive numerical results, we find that only one spin channel of EOPs is
dominated for each of the four choices of $(V_{e},V_{h})$. Furthermore, we
find that the EOPs will obtain $k$-dependent phases due to the Rashba SO
coupling. For convenience of discussion, we define $\chi_{k}$=$\arctan(\sin
k_{y}/\sin k_{x})$ and $\tau_{k}$=$\arctan(b_{k}/a_{k})$. As a typical
example, the numerical results of EOPs for $(V_{e},V_{h})$=$(1,1)$ are shown
in Fig. 2. In this case, one can find from Fig. 2 that the component
$\Delta_{\downarrow\Uparrow}(\vec{k})$ in EOP matrix Eq. (\ref{EOP1}) is
dominant, while the amplitudes of the other three components ($\Delta
_{\uparrow\Uparrow},\Delta_{\uparrow\Downarrow},\Delta_{\downarrow\Downarrow}%
$) are negligibly small. With keeping in mind that the $k$-dependent phases of
EOPs are obviously due to the Rashba SO interaction, we have analytically
constructed various possible SO interaction-induced phases in EOPs and turned
to compare these analytic approximate expressions with our exact numerical
results. Table I summarizes the most optimal approximate phases for the four
magnetic configurations. As an illustration, we plot in Fig. 3 our derived
approximate condensate phases for the case of $(V_{e},V_{h})$=$(1,1)$, and
compare them with the exact numerical result shown in Fig. 2. The agreement is
clear.
\begin{figure}[ptb]
\begin{center}
\includegraphics[width=0.6\linewidth]{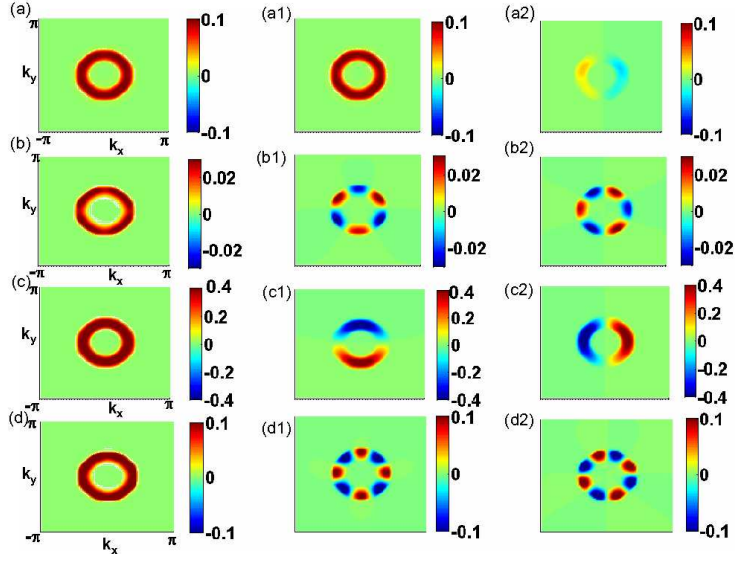}
\end{center}
\caption{(Color online) The left (middle, right) panels respectively show our
calculated magnitudes (real parts, imaginary parts) of the EOPs $\Delta
_{\uparrow\Uparrow}(\vec{k})$, $\Delta_{\uparrow\Downarrow}(\vec{k})$,
$\Delta_{\downarrow\Uparrow}(\vec{k})$, and $\Delta_{\downarrow\Downarrow
}(\vec{k})$, at a typical setup of magnetization parameters ($V_{e},V_{h}%
$)=($1,1$).}%
\end{figure}
\begin{figure}[ptbptb]
\begin{center}
\includegraphics[width=0.6\linewidth]{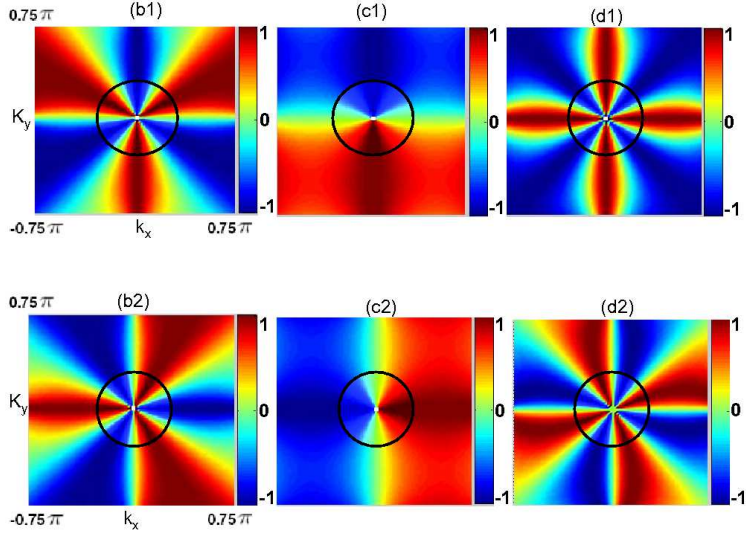}
\end{center}
\caption{(Color online) (from left to right) Phases factors of the EOPs
$\Delta_{\uparrow\Downarrow}(\vec{k})$, $\Delta_{\downarrow\Uparrow}(\vec{k}%
)$, and $\Delta_{\downarrow\Downarrow}(\vec{k})$ that are listed in Table I
with ($V_{e},V_{h}$)=($1,1$). The imaginary part of the component
$\Delta_{\uparrow\Uparrow}(\vec{k})$ is negligibly small (see Fig. 2(a2)).
Here the upper and lower panels respectively plot the real and imaginary parts
of these three phase factors. The black cirques denote Fermi surface. }%
\end{figure}%
\[
\overset{\text{TABLE I. Dominant spin channel and approximate analytical EOP
phase factors for different magnetization configurations}}{%
\begin{tabular}
[c]{ccc}\hline\hline
($V_{e},V_{h}$) & dominant EOP component & phases of ($\Delta_{\uparrow
\Uparrow},\Delta_{\uparrow\Downarrow},\Delta_{\downarrow\Uparrow}%
,\Delta_{\downarrow\Downarrow}$)\\\hline
($1,1$) & $\Delta_{\downarrow\Uparrow}(\vec{k})$ & $(1,-ie^{i\tau_{k}%
},ie^{i\chi_{k}},e^{i(\chi_{k}+\tau_{k})})$\\
($-1,1$) & $\Delta_{\uparrow\Uparrow}(\vec{k})$ & $(ie^{-i\chi_{k}}%
,e^{-i\chi_{k}+i\tau_{k}},-1,ie^{i\tau_{k}})$\\
($-1,-1$) & $\Delta_{\uparrow\Downarrow}(\vec{k})$ & $(e^{-i(\chi_{k}+\tau
_{k})},-ie^{-i\chi_{k}},ie^{-i\tau_{k}},1)$\\
($1,-1$) & $\Delta_{\downarrow\Downarrow}(\vec{k})$ & $(-ie^{i\tau_{k}%
},-1,e^{i\chi_{k}-i\tau_{k}},-ie^{i\chi_{k}})$\\\hline
\end{tabular}
\ \ \ \ \ \ \ \ \ \ \ \ \ \ \ \ \ }\ \
\]

With the help of Table I, we expect that the $k$-dependent phases in the EOPs
may lead to the nontrivially chiral topological orders. For instance, let us
consider the case of $(V_{e},V_{h})$=$(1,1)$. In the continuum limit,
$e^{i\chi_{k}}\mathtt{\sim}\frac{k_{x}+ik_{y}}{k}$, and thus $\Delta
_{\downarrow\Uparrow}(\vec{k})\mathtt{\sim}i|\Delta_{\downarrow\Uparrow}%
(\vec{k})|\frac{k_{x}+ik_{y}}{k}$. That means the ($p_{x}\mathtt{+}ip_{y}%
$)-like pairing emerges.

Moreover, an explicit picture of chiral TEI can be well understood in the
two-band approximation. To reveal this fact, first the non-interacting part in
the total Hamiltonian is rewritten in the single-particle eigenstate space as%

\begin{equation}
\mathcal{H}_{0}=\underset{\vec{k},s}{%
%TCIMACRO{\dsum }%
%BeginExpansion
{\displaystyle\sum}
%EndExpansion
}E_{es}(\vec{k})\psi_{es}^{\dagger}(\vec{k})\psi_{es}(\vec{k})+E_{hs}(\vec
{k})\psi_{hs}(-\vec{k})\psi_{hs}^{\dagger}(-\vec{k}),\label{Han_diag}%
\end{equation}
where $E_{es}$=$\zeta_{\vec{k}}^{(e)}\mathtt{-}\mu_{e}\mathtt{+}s\sqrt
{\alpha^{2}(\sin^{2}k_{x}\text{+}\sin^{2}k_{y})\text{+}V_{e}^{2}}$ and
$E_{hs}$=$-\zeta_{-\vec{k}}^{(h)}$+$\mu_{h}\mathtt{+}s\sqrt{\beta^{2}%
(a_{k}^{2}+b_{k}^{2})\text{+}V_{h}^{2}}$ ($s$=$+,-$) are respectively electron
and heavy-hole band energies, and $\psi_{ps}$ denotes the relevant
annihilation field operators. Here the single-particle eigenstates are given
by%
\begin{align}
\varphi_{e-}(\vec{k}) &  =e^{i\theta_{k}}\left[  -if_{+}(k)e^{-i\chi_{k}%
},f_{-}(k),0,0\right]  ^{T},\nonumber\\
\varphi_{e+}(\vec{k}) &  =e^{i\theta_{k}}\left[  f_{-}(k),-if_{+}%
(k)e^{i\chi_{k}},0,0\right]  ^{T},\nonumber\\
\varphi_{h-}(\vec{k}) &  =e^{i\vartheta_{k}}\left[  0,0,ig_{+}(k)e^{i\tau_{k}%
},g_{-}(k)\right]  ^{T},\nonumber\\
\varphi_{h+}(\vec{k}) &  =e^{i\vartheta_{k}}\left[  0,0,g_{-}(k),ig_{+}%
(k)e^{-i\tau_{k}}\right]  ^{T},\label{Ham_wf}%
\end{align}
where $f_{\pm}(k)$=$\frac{\alpha\sqrt{\sin^{2}k_{x}+\sin^{2}k_{y}}}%
{\sqrt{\alpha^{2}(\sin^{2}k_{x}+\sin^{2}k_{y})+(\sqrt{\alpha^{2}(\sin^{2}%
k_{x}+\sin^{2}k_{y})+V_{e}^{2}}\pm V_{e})^{2}}}$ and $g_{\pm}(k)$=$\frac
{\beta\sqrt{a_{k}^{2}+b_{k}^{2}}}{\sqrt{\beta^{2}(a_{k}^{2}+b_{k}^{2}%
)+(\sqrt{\beta^{2}(a_{k}^{2}+b_{k}^{2})+V_{h}^{2}}\pm V_{h})^{2}}}$. Note that
$\theta_{k}$ and $\vartheta_{k}$ are $k$-dependent phases and are in principal
determined, during exciton formation, by exactly solving the ground state of
the system through our above self-consistent calculation. The single-particle
bands $E_{p\pm}(\vec{k})$ are shown in Fig. 1(b) (solid curves), from which it
is easy to find that the excitons are preferably formed between the lower
electron band $E_{e-}$ and the upper hole band $E_{h+}$. Moreover, the pairing
relates to the Fermi surface of the bilayer system. With the values of the
tunable parameters shown in the caption of the Fig. 1, the band $E_{e-}$ and
band $E_{h+}$ have the nearly perfect nesting Fermi surface with the Fermi
energy $E_{F}$ being nearly zero, (namely, $\mu_{p}$=$-4t_{p}$). Hence, we can
deal with pairing in BCS picture in this situation. Now, we consider the
electron-hole interaction part in Eq. (\ref{Ham1_k}) in terms of the filled
electron band $E_{e-}$ and hole band $E_{h+}$. In order to obtain an explicit
picture, we use a rough approximation by assuming a short-range interaction
potential $U^{(eh)}(q)$=$U\delta(q)$. Then, after mean-field treatment, the
resultant two-band Hamiltonian for our exciton system is given by%
\begin{align}
\bar{H}_{MF} &  \approx\underset{\vec{k}}{%
%TCIMACRO{\dsum }%
%BeginExpansion
{\displaystyle\sum}
%EndExpansion
}E_{e-}(\vec{k})\psi_{e-}^{\dagger}(\vec{k})\psi_{e-}(\vec{k})+\underset
{\vec{k}}{%
%TCIMACRO{\dsum }%
%BeginExpansion
{\displaystyle\sum}
%EndExpansion
}E_{h+}(\vec{k})\psi_{h+}(-\vec{k})\psi_{h+}^{\dagger}(-\vec{k})\nonumber\\
&  -\frac{1}{2}\underset{\vec{k}}{%
%TCIMACRO{\dsum }%
%BeginExpansion
{\displaystyle\sum}
%EndExpansion
}(\bar{\Delta}(\vec{k})\psi_{e-}^{\dagger}(\vec{k})\psi_{h+}^{\dagger}%
(-\vec{k})+\text{H.c.})+\frac{1}{2}\underset{\vec{k}}{%
%TCIMACRO{\dsum }%
%BeginExpansion
{\displaystyle\sum}
%EndExpansion
}\frac{\left\vert \bar{\Delta}(\vec{k})\right\vert ^{2}}{U},\label{Ham_er}%
\end{align}
where $\bar{\Delta}(\vec{k})$=$\underset{s,s^{\prime}}{%
%TCIMACRO{\dsum }%
%BeginExpansion
{\displaystyle\sum}
%EndExpansion
}Uf_{s}^{2}(k)g_{s^{\prime}}^{2}(k)\left\langle \psi_{h+}(-\vec{k})\psi
_{e-}(\vec{k})\right\rangle $ ($s,s^{\prime}$=$\pm$). The straightforward
calculation can prove $\underset{s,s^{\prime}}{%
%TCIMACRO{\dsum }%
%BeginExpansion
{\displaystyle\sum}
%EndExpansion
}f_{s}^{2}(k)g_{s^{\prime}}^{2}(k)\mathtt{\sim}1$ near the Fermi wave vector
$k_{F}$. So $\bar{\Delta}(\vec{k})\mathtt{\approx}\Delta_{0}$ is almost
$k$-independent and only nonzero around $k_{F}$ in BCS-type picture. In
practice we can introduce a factor $\gamma(\vec{k})$=$e^{-c(k-k_{F})^{2}%
}e^{i\omega}$ ($c$ and $\omega$ are real constants) to fit our exact
multi-band self-consistent numerical results (say, Fig. 2) in the whole BZ.
The gaped energy spectrum of $\bar{H}_{MF}$ is shown in Fig. 1(b) (dashed
lines). Now, in the two-band approximation, the EOPs in Eq. (\ref{EOP1}) have
the expressions as follows:%
\begin{equation}
\mathbf{\Delta}(\vec{k})=\frac{\Delta_{0}e^{-c(k-k_{F})^{2}}e^{i(\theta
_{k}-\vartheta_{k}+\omega)}}{2}\left[
\begin{array}
[c]{cc}%
if_{+}(k)g_{-}(k)e^{-i\chi_{k}} & f_{+}(k)g_{+}(k)e^{-i\chi_{k}+i\tau_{k}}\\
-f_{-}(k)g_{-}(k) & if_{-}(k)g_{+}(k)e^{i\tau_{k}}%
\end{array}
\right]  ,\label{EOP3}%
\end{equation}
where the phases $e^{i(\theta_{k}-\vartheta_{k}+\omega)}$ are confirmed
through our self-consistent calculation. It turns out that Equation
(\ref{EOP3}) gives a nice description of the numerical results.

\section{ Chiral topological order}

In the presence of exchange fields ($V_{e},V_{h}$) induced by the
ferromagnetic films, the TRS of the system is broken. No less than the AQHE,
the nonzero TKNN number can undoubtedly characterize the topological nature of
the system if a stable bulk gap separates the ground state and excited states.
That means the topological property of the system will not be changed without
bulk gap closing in spite of adiabatically deforming
%TCIMACRO{\TEXTsymbol{\vert}}%
%BeginExpansion
$\vert$%
%EndExpansion
$\Delta_{\sigma\sigma^{\prime}}(\vec{k})|$ at the given exchange fields.
Hence, $\gamma(\vec{k})$ in Eq. (\ref{EOP3}) is inessential for the system's
topological property. Moreover, only the dominant component of EOPs decides
the system's topological property at the given ($V_{e},V_{h},$). The
straightforward calculation of $I_{TKNN}$ in Eq. (\ref{chern}) can prove the
above two arguments.

In the following discussion, we use Eq. (\ref{EOP3}) to consider the system's
topological properties. In general, in the spin-dependent Nambu space
$(e_{\vec{k}\mathbf{\uparrow}}$ $e_{\vec{k}\mathbf{\downarrow}}$ $h_{-\vec
{k}\mathbf{\Uparrow}}^{\dagger}$ $h_{-\vec{k}\mathbf{\Downarrow}}^{\dagger})$,
the EOPs in different spin channels are affected by the effective exchange
fields. Additionally, the strong Rashba SO interaction flaws the spin
polarization of the carries along the $z$ direction. The total effect leads
the factors $f_{\pm}(k)g_{\pm}(k)$ to emerge in different spin channels of
EOPs, and which decide the dominant one at given ($V_{e},V_{h},$). For
convenience of the following discussion, we use ($\Delta_{0}^{uu},\Delta
_{0}^{ud},\Delta_{0}^{du},\Delta_{0}^{dd}$) to denote $\Delta_{0}$%
($f_{+}(k)g_{-}(k),f_{+}(k)g_{+}(k),f_{-}(k)g_{-}(k),f_{-}(k)g_{+}(k)$). The
topological nature of the ground state $|u_{0}(\vec{k})\rangle$ can be
charactered by non-zero $I_{TKNN}$, which reads%
\begin{equation}
I_{TKNN}=-\frac{1}{2\pi}\int_{BZ}d^{2}k\Omega_{0}(\vec{k}), \label{chern}%
\end{equation}
where $\Omega_{0}(\vec{k})$=$-2\operatorname{Im}\left\langle \frac{\partial
u_{0}}{\partial k_{x}}\right\vert \left.  \frac{\partial u_{0}}{\partial
k_{y}}\right\rangle $ is the ground-state Berry curvature in BZ. The results
are summarized in Table II, which definitely shows chiral topological order
with its winding behavior depending on the choice of exchange-field
parameters. From the bulk-edge correspondence, the nontrivial bulk topological
number implies gapless edge states in the system with finite size.
\[
\overset{\text{TABLE II. The TKNN numbers for effective exchange fields and
corresponding EOP amplitudes}}{%
\begin{tabular}
[c]{cccc}\hline\hline
($V_{e},V_{h}$) & $\Delta_{0}$ & ($\Delta_{0}^{uu},\Delta_{0}^{ud},\Delta
_{0}^{du},\Delta_{0}^{dd}$) & $I_{TKNN}$\\\hline
($1,1$) & $0.5$ & $0.5$($0,0,1,0$) & $1$\\
($-1,1$) & $0.5$ & $0.5$($1,0,0,0$) & $-1$\\
($-1,-1$) & $0.5$ & $0.5$($0,1,0,0$) & $-1$\\
($1,-1$) & $0.5$ & $0.5$($0,0,0,1$) & $1$\\\hline
\end{tabular}
\ \ \ \ \ \ \ \ \ \ \ \ }\ \
\]
\ \

\begin{figure}[ptb]
\begin{center}
\includegraphics[width=0.8\linewidth]{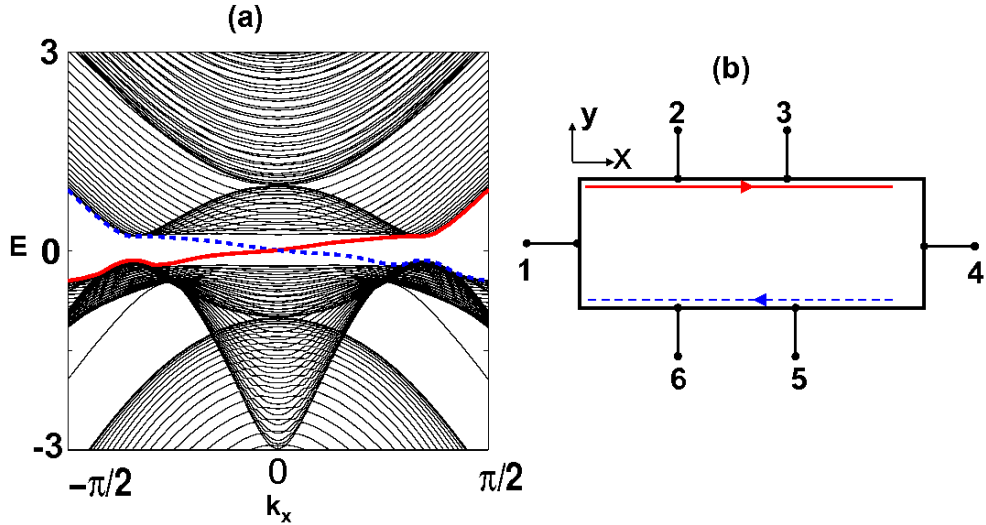}
\end{center}
\caption{(Color online) (a) The energy spectrum of the bilayer square-lattice
system with two edges at the $y$ direction. $k_{x}$ denotes the momentum in
the $x$ direction. The magnetization parameters are set at ($V_{e},V_{h}%
$)=($1,1$). The red-solid and blue-dashed lines denote the edge states
locating on different edges. (b), Six-terminal Hall bar for detection of the
edge states. The red-solid and blue-dashed lines with arrows represent the
edge modes propagating in opposite direction. }%
\end{figure}

In order to confirm the existence of the gapless edge states, we assume that
the square-lattice system has two edges in $y$ direction and is boundless in
$x$ direction. Correspondingly, we choose open boundary condition in $y$
direction and periodic boundary condition in $x$ direction of the lattice
Hamiltonian in Eq (\ref{Ham1}) in mean-field approximation. The calculated
energy spectrum at a typical case of ($V_{e},V_{h}$)=($1,1$) is illustrated in
Fig. 4 (a). The red-solid and blue-dashed lines correspond to the different
edge states with contrary chirality. It is easy to find that the number of the
gapless edge states is consistent with the bulk theory characterized by
$I_{TKNN}$.

\section{Transport property of edge states}

The nontrivial transport phenomena can be predicted due to the emergence of
the edge states in our system. From Fig. 4 (a), \ we can find that the edge
sates in different chiral topological order propagate on each boundary with
opposite velocities and can be described by the following Hamiltonian%

\begin{equation}
H_{edge}^{\eta}=\pm\underset{k_{x}\geq0}{\sum}\lambda_{\eta}v_{F}k_{x}%
\gamma_{\eta}^{\dagger}(k_{x},y)\gamma_{\eta}(k_{x},y), \label{edge Ham}%
\end{equation}
where $\pm$ represents different edges and $\eta$=1,...,4 labels four
different kinds of magnetic configurations, namely, $\lambda_{1}$=$\lambda
_{4}$=$1$ and $\lambda_{2}$=$\lambda_{3}$=$-1,v_{F}$ is the Fermi velocity and
$k_{x}$ is the momentum measured from the Fermi surface. The quasiparticle
operators for case ($V_{e},V_{h}$)=($1,1$) read
\begin{equation}
\gamma_{1}(k_{x},y)=u_{1}(k_{x},y)e_{\uparrow}(y)+v_{1}(k_{x},y)h_{\Uparrow
}^{\dagger}(y). \label{quasi oper}%
\end{equation}
The other cases have the similar forms. Due to the missing of the PHS, the
quasiparticles are not Majorana fermions.

The edge states in the AQHE systems can be usually detected through the Hall
conductance responding to the external electromagnetic field \cite{McEuen}%
\cite{Wang}. However, the edge states in our system are excitons which are
charge neutral. A simple approach is to use thermal transport measurement
which is often used to judge the pair properties in high-$T_{c}$
superconductors \cite{Hill}\cite{Smith}. The six-terminal Hall bar showed in
Fig. 4 (b) for detecting the edge states of quantum (spin) Hall effect can be
used to detect the thermal conductance. The same setup has been used by Sato
\emph{et.al }\cite{Sato}\emph{ }to detect the edge state in topological
superconductor. We give the similar considerations with that in Ref.
\cite{Sato} as follows. The temperature must be sufficiently lower than the
exciton gap ($T\mathtt{\ll}\Delta_{0}$) in order to suppress the contributions
from the fermionic excitations (electrons and holes) in the bulk and bosonic
(phonons) excitations. The thermal conductance is defined by $G(T)$%
=$I_{14}(T)/(\Delta T)_{14}$, where $I_{ij}(T)$ is a thermal current between
contacts $i$ and $j$, and $(\Delta T)_{ij}$ is the temperature difference
between these contacts. In the low temperature limit, the $T$-dependence of
$G(T)$ have three origins: the linear law $\mathtt{\varpropto}T$ from edge
states for phase $\eta$, the exponentially low $\mathtt{\sim}e^{-\Delta_{0}%
/T}$ from bulk quasiparticles and the power law $\mathtt{\varpropto}T^{3}$
from phonons. Furthermore, in analogy with the quantum spin Hall current
discussed in Ref. \cite{Bernevig}, there is no temperature difference between
contacts 2 and 3 (5 and 6) because the edge current is dissipationless.

\section{conclusion}

In conclusion, we have presented a scheme to realize the chiral topological
excitonic insulator in the double quantum wells adjacent to two ferromagnetic
films. We have predicted different topologically nontrivial orders emergent
along with changes in the magnetization orientations in the ferromagnetic
films. The topologically nontrivial orders can be characterized by the chiral
topological numbers defined with TKNN numbers in bulk system or chiral edge
states in edged system. Furthermore, we have given an experimental scheme to
detect the excitonic gapless edge states.

\begin{acknowledgments}
This work was supported by NSFC under Grants No. 90921003, No. 10574150 and
No. 60776063, and by the National Basic Research Program of China (973
Program) under Grants No. 2009CB929103, and by a grant of the China Academy of
Engineering and Physics.
\end{acknowledgments}


\begin{thebibliography}{99}                                                                                               %


\bibitem {Haldane}F. D. M. Haldane, Phys. Rev. Lett. \textbf{61}, 2015 (1988).

\bibitem {TKNN}D. J. Thouless, M. Kohmoto, M. P. Nightingale and M. den Nijs,
Phys. Rev. Lett. \textbf{49}, 405 (1982).

\bibitem {Halperin}B. I. Halperin, Phys.Rev. B \textbf{25}, 2185 (1982).

\bibitem {Hatsugai}Y. Hatsugai, Phys. Rev. Lett. \textbf{71}, 3697 (1993)

\bibitem {Liu}C. X. Liu, X. L. Qi, X. Dai, Z. Fang, and S. C. Zhang, Phys.
Rev. Lett. \textbf{101}, 146802 (2008).

\bibitem {Read}N. Read and D. Green, Phys. Rev. B \textbf{61},10267 (2000)

\bibitem {Nitta}J. Nitta, T. Akazaki, and H. Takayanagi, and T. Enoki, Phys.
Rev. Lett. \textbf{78}, 1335 (1997).

\bibitem {Lu}J. P. Lu, J. B. Yau, S. P. Shukla, M. Shayegan, L. Wissinger, U.
R\"{o}ssler, and R. Winkler, Phys. Rev. Lett. \textbf{81}, 1282 (1998).

\bibitem {Grundler}D. Grundler, Phys. Rev. Lett. \textbf{84}, 6074 (2000).

\bibitem {Papadakis}S. J. Papadakis, E. P. De Poortere, H. C. Manoharan, M.
Shayegan, R. Winkler, Science \textbf{283}, 2056 (1999)

\bibitem {Gor'kov}Lev P. Gor'kov and E. I. Rashba, Phys. Rev. Lett.
\textbf{87}, 037004 (2001).

\bibitem {Engels}G. Engels, J. Lange, Th. Sch\"{a}pers, and H. L\"{u}th, Phys.
Rev. B \textbf{55}, R1958 (1997).

\bibitem {Winkler}R. Winkler, Phys. Rev. B \textbf{62}, 4245 (2000).

\bibitem {Winkler1}R. Winkler, H. Noh, E. Tutuc, and M. Shayegan, Phys. Rev. B
\textbf{65}, 155303 (2002).

\bibitem {Yang}C. L. Yang, H. T. He, L. Ding, L. J. Cui, Y. P. Zeng, J. N.
Wang, and W. K. Ge, Phys. Rev. Lett. \textbf{96}, 186605 (2006).

\bibitem {McEuen}P. L. McEuen, A. Szafer, C. A. Richter, B. W. Alphenaar, J.
K. Jain, A. D. Stone, R. G. Wheeler, and R. N. Sacks, Phys. Rev. Lett.
\textbf{64}, 2062 (1990).

\bibitem {Wang}J. K. Wang and V. J. Goldman, Phys. Rev. Lett. \textbf{67}, 749 (1991).

\bibitem {Hill}R. W. Hill, Cyril Proust, Louis Taillefer, P. Fournier, and R.
L. Greene, Nature \textbf{414}, 711 (2001).

\bibitem {Smith}M. F. Smith, Johnpierre Paglione, and M. B. Walker and Louis
Taillefer, Phys. Rev. B \textbf{71}, 014506 (2005).

\bibitem {Sato}M. Sato and S. Fujimoto, Phys. Rev. B \textbf{79}, 094504 (2009).

\bibitem {Bernevig}B. A. Bernevig, T. L. Hughes, and S. C. Zhang, Science 314,
1757 (2006).
\end{thebibliography}
\end{document}